\documentclass{article}
\usepackage{amsmath}
\usepackage{graphicx}
\usepackage{amsfonts}

\title{$\Delta-$string - a hybrid between wormhole and string}
\author{V. Dzhunushaliev
\thanks{E-mail: dzhun@hotmail.kg}}

\date{}

\begin{document}
\maketitle

\begin{center}
\textit{Dept. Phys. and Microel. Engineer., KRSU, Bishkek, \\
Kievskaya Str. 44, 720000, Kyrgyz Republic}
\end{center}

\begin{abstract}
The flux tube solutions in 5D Kaluza-Klein theory can be considered 
as a string-like object - $\Delta-$string. The initial 5D metric can be 
reduced to some inner degrees of freedom living on the $\Delta-$string. 
The propagation of electromagnetic waves through the $\Delta-$string 
is considered. It is shown that the difference between $\Delta$ and 
ordinary strings are connected with the fact that for the $\Delta-$string 
such limitations as critical dimensions are missing. 
\end{abstract}

\section{Introduction}

The difference between point-like particles and strings on the one hand, 
and Einstein's point of view on an inner structure of matter on the other 
hand is that according to Einstein everything must have an inner structure. 
Even more : at the origin of matter should be vacuum. In string theory a 
string is a vibrating 1-dimensional object and the string has many different 
harmonics of vibration, and in this context different elementary particles 
are interpreted as different harmonics of the string. The string degrees of 
freedom are the coordinates of string points in an ambient space. 
\par
\begin{figure}[h]
  \begin{minipage}[t]{.45\linewidth}
  \begin{center}
    \fbox{
    \includegraphics[height=5cm,width=6cm]{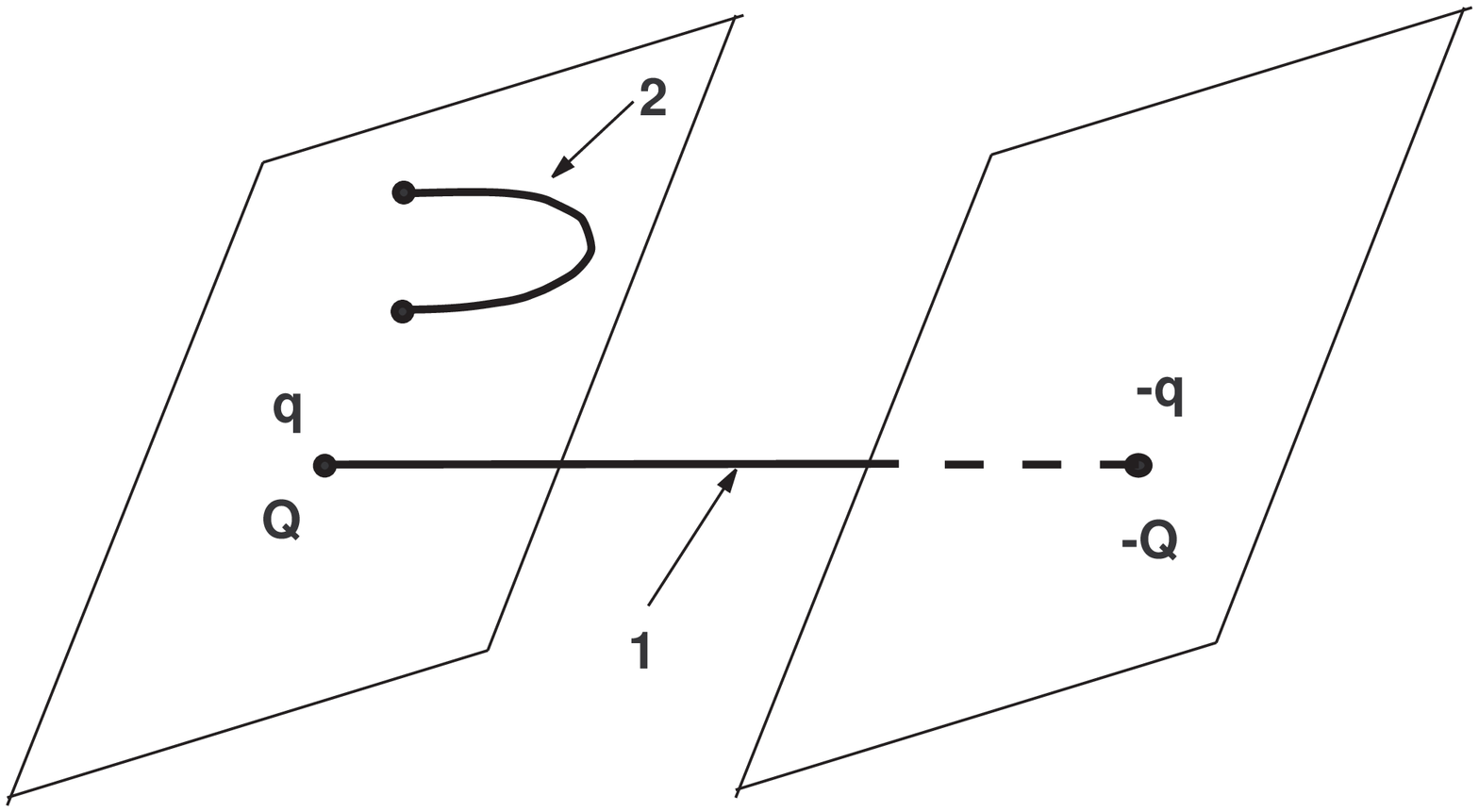}}
    \caption{The gravitational flux tube attached to 
    two Universes \textbf{(1)} or to remote parts of a single Universe
    \textbf{(2)}. 
    The cross section is in order of the Planck length and the tube 
    length can be arbitrary long. Such construction is similar 
    to the string attached to D-brane(s).}
    \label{fig1}
  \end{center}  
  \end{minipage}\hfill
  \begin{minipage}[t]{.45\linewidth}
  \begin{center}
    \fbox{
    \includegraphics[height=5cm,width=6cm]{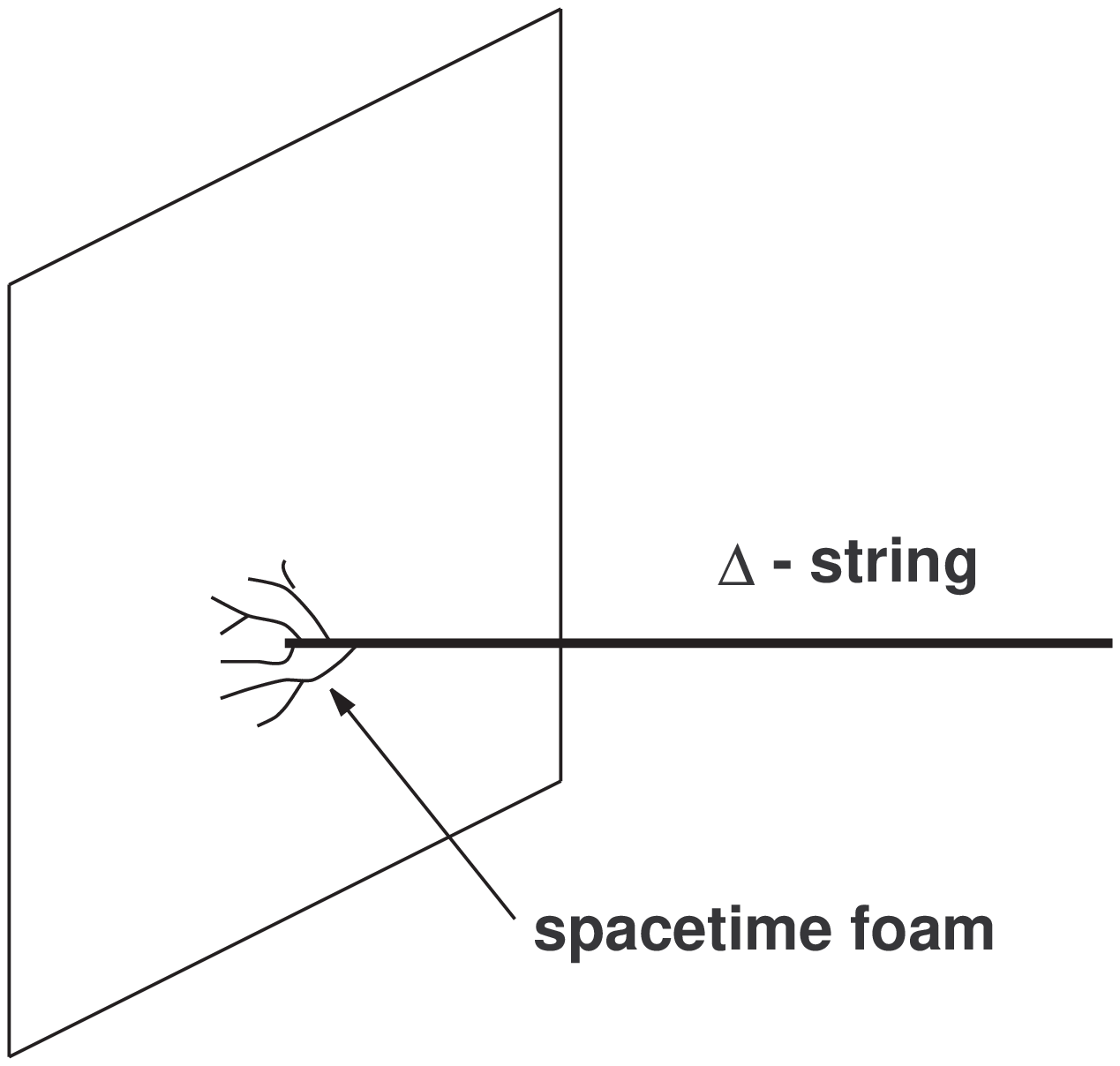}}
    \caption{$\Delta-$string is the gravitational flux tube surrounded 
    with the handles of spacetime foam at the ends. }
    \label{fig2}
  \end{center}  
  \end{minipage} 
\end{figure}

 
In this paper we would like to consider the situation when the string has 
an inner structure. The question in this situation is : what will be changed 
in this situation ? Definitely we can say that in this situation the string 
is an object which effectively arise from a field theory and such object has 
inner degrees of freedom which are not connected with an external space. As a 
model of such kind of string-like object we will consider gravitational flux 
tubes. These tubes are the solutions in 5D Kaluza-Klein gravity \cite{dzhsin1} 
filled with electric $E$ and magnetic $H$ fields. If $E=H$ we have an infinite 
tube, if $E \approx H$ (but $E>H$) the length of the tube can be arbitrary 
long and the cross section can be $\approx l_{Pl}$ ($l_{Pl}$ is the Planck length) 
\cite{dzh2}. Such flux tube can be considered as a string attached to two 
Universes or to remote parts of a single Universe. 
For the observer in the outer Universe the attachment points looks like to 
point-like electric and magnetic charges (see Fig. \ref{fig1}). 
\par 
We have to note that similar construction was presented in Ref. 
\cite{guendelman} : the matching of two remote regions was done using 4D infinite 
flux tube which is the Levi-Civita - Bertotti - Robinson solution \cite{Levi-Civita}, 
\cite{Bertotti} filled with the electric and magnetic fields. 

\section{Gravitational flux tube}

In this section we will describe the gravitational flux tube and why it can be 
considered as a $\Delta-$string. Let us consider the following 5D metric 
\begin{equation} 
\begin{split}
ds^2 = e^{2\nu (r)}dt^{2} - e^{2\psi (r) - 2\nu (r)}
\left [d\chi +  \omega (r)dt + Q \cos \theta d\varphi \right ]^2  \\
- dr^{2} - a(r)(d\theta ^{2} + 
\sin ^{2}\theta  d\varphi ^2),
\label{sec1-10}
\end{split}
\end{equation}
where $\chi $ is the 5th extra coordinate; 
$r,\theta ,\varphi$ are $3D$  spherical-polar coordinates; 
$Q$ is the magnetic charge. The solution of 5D Kaluza-Klein equations 
depends on the relation $\delta = 1 - H/E$ between electric $E$ 
and magnetic $H$ fields. 
If $\delta = 0$ we have an infinite flux tube filled 
with electric and magnetic fields \cite{dzhsin1}
\begin{eqnarray} 
a_0 = \frac{q^2}{2} = const, 
\label{sec1-20}\\
e^{\psi} = e^{\nu} = \cosh\frac{r\sqrt{2}}{q},
\label{sec1-30}\\
\omega = \sqrt{2}\sinh\frac{r\sqrt{2}}{q} 
\label{sec1-40}
\end{eqnarray} 
here $q$ is the electric charge. 
If $0 < \delta \ll 1$ the part of spacetime located between two hypersurfaces 
$ds^2(r = \pm r_H) = 0$ 
\footnote{at these points $\exp (-2\nu (r = \pm r_H)) = 0$} 
is a finite flux tube filled with electric and magnetic fields. 
In both cases the cross sectional sizes can be arbitrary but we choose its in the 
Planck region ($\approx l^2_{Pl}$). This condition is very important for 
the idea presented here : the gravitational flux tube can be considered as the 
$\Delta-$string attached to two different Universes (or to remote parts of a 
single Universe). From the physical point of view this is 1D object as the 
Planck length is a minimal length in the physical world. 
\par 
Now our strategy is to consider small perturbations of 5D metric. Generally 
speaking they are 5D gravitational waves on the gravitational flux tube (or on 
the string language - vibrations of $\Delta-$string). In the general case 
5D metric is 
\begin{equation}
  ds^2 = g_{\mu\nu} dx^\mu dx^\nu - \phi^2
  \left(
  d\chi + A_\mu dx^\mu
  \right)^2
\label{sec1-50}
\end{equation}
here $g_{\mu\nu}$ is the 4D metric; $\mu , \nu = 0,1,2,3$; $A_\mu$ is the 
electromagnetic potential; $\phi$ is the scalar field. The corresponding 5D 
Kaluza-Klein's equations are  (for the reference see, for example, \cite{wesson}) 
\begin{eqnarray}
R_{\mu \nu} - \frac{1}{2} g_{\mu \nu} R & = & 
-\frac{\phi^2}{2} T_{\mu \nu} - \frac{1}{\phi} 
\left [
\nabla _\mu \left (
            \partial _\nu \phi
            \right ) - g_{\mu \nu} \Box \phi 
\right ] ,
\label{sec1-60}\\
\nabla _\nu F^{\mu \nu} & = & -3 \frac{\partial _\nu \phi}{\phi} F^{\mu \nu} ,
\label{sec1-70}\\
\Box \phi & = & - \frac{\phi^3}{4} F^{\alpha \beta} F_{\alpha \beta} 
\label{sec1-80}
\end{eqnarray}
where $R_{\mu \nu}$ is the 4D Ricci tensor; 
$F_{\mu \nu} = \partial_\mu A_\nu - \partial_\nu A_\mu$ is the 
4D Maxwell tensor and $T_{\mu \nu}$ is the energy-momentum tensor 
for the electromagnetic field. 
\par
We will consider only $\delta A_\mu$ perturbations, $\delta g_{\mu\nu}$ 
and $\delta \phi$ degrees of freedom are frozen. For this approximation 
we have equation 
\begin{equation}
  \nabla_\nu \delta F^{\mu\nu} = 0 .
\label{sec1-90}
\end{equation}
We introduce only one small perturbation in the electromagnetic potential 
\begin{equation}
  \delta A_\theta = f(t,r) 
\label{sec1-100}
\end{equation}
since for the background metric $\phi = const$. 
Generally speaking $A_\theta$-component should  have some dependence on 
the $\theta$-angle. But the cross section of the $\Delta-$string is in Planck region 
and consequently the points with different $\theta$ and $\varphi$ 
($r$ = const) physically are not distinguishable. Therefore all physical 
quantities on the $\Delta-$string should be averaged over polar angles 
$\theta$ and $\varphi$. It means that the $\delta A_\theta$ in 
Eq. \eqref{sec1-100} is averaged quantity. 
\par 
After this (very essential) remark we have the following wave equation 
for the function $f(t,r)$ 
\begin{equation}
  \partial_{tt} f(t,x) - \cosh x \partial_x 
  \biggl (
  \cosh x \partial_x f(t,x)
  \biggl ) = 0
\label{sec1-110}
\end{equation}
here we have introduced the dimensionless variables $t/\sqrt{a_0} \rightarrow t$ 
and $r/\sqrt{a_0} \rightarrow x$. The solution is 
\begin{equation}
  f(t,x) = f_0 F\left(t - 2 \arctan e^x \right) + 
  f_1 F\left(t + 2 \arctan e^x \right)
\label{sec1-120}
\end{equation}
here $f_{0,1}$ are some constants and $F$ and $H$ are arbitrary functions. 
This solution has more suitable form if we introduce new coordinate 
$y = 2 \arctan e^x$. Then 
\begin{equation}
  f(t,y) = f_0 F(t-y) + f_1 H(t + y).
\label{sec1-130}
\end{equation}
The metric is 
\begin{equation}
  ds^2 = \frac{a_0}{\sin^2y} dt^2 - \frac{dy^2}{\sin^2y} - 
  a_0\left( d\theta^2 + \sin^2\theta d\varphi^2 \right) - 
  \left(
  d\chi + \omega dt + Q \cos \theta d\varphi
  \right)^2 .
\label{sec1-140}
\end{equation}
Thus the simplest solution is electromagnetic waves moving in both 
directions along the $\Delta-$string. 

\section{The comparison with string theory}

Now we want to compare this situation with the situation in string theory. 
How is the difference between the result presented here 
(Eq. \eqref{sec1-110}, \eqref{sec1-120}) 
and the string oscillation in the ordinary string theory ? The action for 
bosonic string is 
\begin{equation}
  S = - \frac{T}{2} \int d^2 \sigma \sqrt{h} h^{ab} 
  \partial_a \mathcal X^\mu \partial_b \mathcal X_\mu 
\label{sec2-10}
\end{equation}
here $\sigma^a = \{ \sigma , \tau \}$ is the coordinates on the world sheet of 
string; $\mathcal X^\mu$ are the string coordinates in the ambient spacetime; 
$h_{ab}$ is the metric on the world sheet. The variation with 
respect to $\mathcal X^\mu$ give us the usual 2D wave equation 
\begin{equation}
  \Box \mathcal X^\mu \equiv 
  \left(
  \frac{\partial^2}{\partial \sigma ^2} - 
  \frac{\partial^2}{\partial \tau ^2} 
  \right) \mathcal X^\mu = 0 
\label{sec2-20}
\end{equation}
which is similar to Eq. \eqref{sec1-110}. The difference is that the variation 
of the action \eqref{sec2-10} with respect to the metric $h^{ab}$ gives us 
some constraints equations in string theory 
\begin{equation}
  T_{ab} = \partial_a \mathcal X^\mu \partial_b \mathcal X_\mu - 
  \frac{1}{2} h_{ab} h^{cd} 
  \partial_c \mathcal X^\mu \partial_d \mathcal X_\mu = 0 
\label{sec2-30}
\end{equation}
but for $\Delta-$string analogous variation gives the dynamical equation 
for 2D metric $h_{ab}$. 
\par 
The more detailed description is the following. The topology of 5D 
Kaluza-Klein spacetime is $M^2 \times S^2 \times S^1$ 
where $M^2$ is the 2D space-time spanned on the time and longitudinal 
coordinate $r$; $S^2$ is the cross section of the flux tube solution and it is 
spanned on the ordinary spherical coordinates $\theta$ and $\varphi$; 
$S^1 = U(1)$ is the Abelian gauge group which in this consideration is 
the 5th dimension. The initial 5D action for 
$\Delta-$string is 
\begin{equation}
  S = \int \sqrt{-\stackrel{(5)}{g}} \stackrel{(5)}{R} 
\label{sec2-31}
\end{equation}
where $\stackrel{(5)}{g}$ is the 5D metric \eqref{sec1-50}; 
$\stackrel{(5)}{R}$ are 5D Ricci scalars. We want to reduce the initial 
5D Lagrangian \eqref{sec2-31} to a 2D Lagrangian (here we will follow to 
Ref. \cite{dzh3}). Our basic assumption is that the sizes of 5th and dimensions 
spanned on polar angles $\theta$ and $\varphi$ is approximately 
$\approx l_{Pl}$.  At first we have usual 5D $\rightarrow$ 4D 
Kaluza-Klein dimensional reduction. Following, for example, to review 
\cite{wesson} we have  
\begin{equation}
\frac{1}{16 \pi \stackrel{(5)}{G}}  \stackrel{(5)}{R} = 
\frac{1}{16 \pi G} \stackrel{(4)}{R} - \frac{1}{4} \phi^2 
F_{\mu \nu} F^{\mu \nu} + \frac{2}{3} 
\frac{\partial_\alpha \phi \partial^\alpha \phi}{\phi^2}
\label{sec2-40}
\end{equation}
where $\stackrel{(5)}{G} = G \int dx^5$ is 5D gravitational constant; 
$G$ is 4D gravitational constant; $\stackrel{(4)}{R}$ is the 4D Ricci scalars. 
The determinants of 5D 
and 4D metrics are connected as $\stackrel{(5)}{g} = \stackrel{(4)}{g}\phi$. 
One of the basis paradigm of quantum gravity is that a minimal 
length in the Nature is the Planck scale. Physically it means that not any 
physical fields depend on the 5th, $\theta$ and $\varphi$ coordinates. 
\par 
The next step is reduction from 4D to 2D. The 4D metric can be expressed as 
\begin{eqnarray}
d\stackrel{(4)}{s^2} & = & g_{\mu \nu} dx^\mu dx^\nu = 
g_{ab}(x^c) dx^a dx^b + 
\nonumber \\
&&\chi(x^c)
\left (
\omega ^{\bar i}_i dy^i + B^{\bar i}_a(x^c) dx^a
\right )
\left (
\omega ^{\bar j}_j dy^j + B_{\bar i a}(x^c) dx^a
\right ) \eta_{\bar i \bar j}
\label{sec2-60}
\end{eqnarray}
where $a,b, c = 0,1$; $x^a$ are the time and longitudinal coordinates; 
$-\left (\omega^{\bar i}_i dy^i \right ) 
\left (\omega ^{\bar j}_j dy^j \right ) \eta_{\bar i \bar j} = dl^2$ 
is the metric 
on the 2D sphere $S^2$; $y^i$ are the coordinates on the 2D sphere $S^2$; 
all physical quantities $g_{ab}, \chi$ and $B_{\bar i a}$ can depend only 
on the physical coordinates $x^a$. Accordingly to Ref. \cite{coq} we have 
the following dimensional reduction to 2 dimensions 
\begin{eqnarray}
\stackrel{(4)}{R} & = & \stackrel{(2)}{R} + R(S^2) - 
\frac{1}{4} \phi^{\bar i}_{ab} \phi^{ab}_{\bar i} - 
\nonumber \\
&&\frac{1}{2} h^{ij}h^{kl}
\left (
D_a h_{ik} D^a h_{jl} + D_a h_{ij} D^a h^{kl}
\right ) - 
\nabla^a 
\left (
h^{ij} D_a h_{ij}
\right )
\label{sec2-70}
\end{eqnarray}
where $\stackrel{(2)}{R}$ is the Ricci scalar of 2D spacetime; 
$D_\mu$ and $\phi^{\bar i}_{ab}$ are, respectively, the covariant derivative 
and the curvature of the principal connection $B^{\bar i}_a$ and 
$R(S^2)$ is the Ricci scalar of the sphere 
$S^2 =  \mathrm{SU(2)/U(1)}$ with 
linear sizes $\approx l_{Pl}$; $h_{ij}$ is the metric on $\mathcal L$ 
\begin{eqnarray}
\mathrm{su}(2) & = & \mathrm{Lie (SU(2))} = \mathrm{u}(1) \oplus \mathcal L ,
\label{sec2-80}\\
\mathrm{u}(1) & = &\mathrm{Lie(U(1))}
\label{sec2-100}
\end{eqnarray}
here $\mathcal L$ is the orthogonal complement of the u(1) algebra in the su(2) 
algebra; the index $i \in \mathcal L$. The metric $h_{ij}$ is 
proportional to the scalar $\chi$ in Eq. \eqref{sec2-30}. 
\par 
The 2D action (which is our goal) is 
\begin{equation}
  S = \int d^2 x^a \phi \left( \det g_{ab} \right) 
  \left( \det {\omega^{\bar i}_i} \right) 
  \left[
  \frac{\stackrel{(2)}{R}}{16 \pi G} - \frac{1}{4} \phi^2 
    \left(
    F_{t \theta} F^{t \theta} + F_{r \theta} F^{r \theta} 
    \right) + \text{other terms} 
  \right] .
\label{sec2-110}
\end{equation}
The second term in the $[\ldots ]$ brackets give us the wave equation 
\eqref{sec1-110} and the most important is that the first term is not total 
derivative in contrast with the situation in ordinary string theory in the 
consequence of the factors $\phi$ and 
$\left( \det {\omega^{\bar i}_i}\right)$. Therefore the variation with respect 
to 2D metric $g_{ab}$ give us some dynamical equations contrary to string theory 
where this variation leads to the constraint equations \eqref{sec2-30}. 
\par 
This remark allows us to say that the $\Delta-$string do not have such peculiarities 
as critical dimensions (D=26 for bosonic string). The reason for this is very 
simple : the comprehending space for $\Delta-$string is so small that it coincides 
with $\Delta-$string. Thus we can suppose that the critical dimensions 
in string theory is connected with the fact that the string curves the 
external space but the back reaction of curved space on the metric of 
the string world sheet is not taken into account. 

\section{Discussion and conclusions}

Finally : the most important difference between $\Delta$ and ordinary strings 
is that in the first case the dynamical equation for 2D metric is replaced 
by the constraint equation for the second case. Physically it means that in the 
second case we ignore the back reaction of string on the metric of world sheet. 
It is supposed that in ordinary string theory this back reaction is zero 
but for the $\Delta-$string this reaction is very big since the string 
coincides with the ambient space. 
\par 
One can say that the $\Delta-$string is a hybrid between the wormhole and 
the string as it is the wormhole-like solution in 5D Kaluza-Klein gravity 
on the one hand and it is approximately 1D object on the other hand. 
\par 
For the outer observer the attachment point of $\Delta-$string to outer 
Universe looks like a distributed electric and magnetic charges as this 
point is spread in the consequence of the appearance of spacetime foam 
handles between $\Delta-$string and the outer Universe. The $\Delta-$string 
is a bridge between two Universes (or remote parts of a single Universe) like 
to wormhole on the one hand and has an arbitrary long throat like to the string. 
The $\Delta-$string can be considered as a model of Wheeler's ``mass without 
mass'' and ``charge without charge''. We have shown that such object can 
transfer the electromagnetic waves from one Universe to another one or from 
one part of Universe to another one. 
\par 
The discussion of electromagnetic waves propagating through the $\Delta-$string 
is not full as we have frozen $\delta g_{\mu\nu}$ and $\delta\phi$ 
perturbations. Evidently $\delta A_\mu$ perturbations will initiate 
$\delta g_{\mu\nu}$ and $\delta\phi$ waves and certainly the back reaction 
takes place. 

\section{Acknowledgment}
I am very grateful to the ISTC grant KR-677 for the financial support.


\begin{thebibliography}{99}

\bibitem{dzhsin1}
V. Dzhunushaliev and D. Singleton, Phys. Rev. \textbf{D59},
064018 (1999).

\bibitem{dzh2}
V. Dzhunushaliev, Class. Quant. Grav., \textbf{19}, 4817 (2002). 

\bibitem{guendelman}
E. I. Guendelman, Gen. Relat. Grav., \textbf{23}, 1415 (1991).

\bibitem{Levi-Civita}{Levi-Civita, T. (1917). {\it Atti Acad.
Naz. Lincei} {\bf 26},519}

\bibitem{Bertotti}{Bertotti, B. (1959). {\it Phys. Rev.}
{\bf 116}, 1331; Robinson, I. (1959). {\it Bull. Akad. Pol.}
{\bf 7}, 351}

\bibitem{dzh3}
V. Dzhunushaliev, ``Strings from Flux Tube Solutions in Kaluza-Klein 
Theory'', gr-qc/0208023, to be published in Phys. Lett. B.

\bibitem{wesson}
J. M. Overduin and P. S. Wesson, Phys. Rept., \textbf{283}, 303 (1997).

\bibitem{coq}
R. Coquereaux and A. Jadczyk, Commun. Math. Phys., \textbf{90}, 79 (1983). 

\end{thebibliography}
\end{document}